%% file: main.tex
\documentclass[sigconf,9pt, nonacm]{acmart}
\usepackage[english]{babel}

\usepackage{blindtext}
\usepackage{tikz}
\usepackage{amsmath}

\usepackage{filecontents}

\usepackage{tikz}
\usepackage{xcolor}
\usepackage{xspace}
\usepackage{tikz}
\usepackage{pifont}
\usepackage[english]{babel}
\usepackage{blindtext}
\usepackage{lipsum}
\usepackage{colortbl} 
\usepackage[ruled,vlined,linesnumbered]{algorithm2e}
\usepackage{booktabs}

\usepackage{filecontents}
\usepackage{xspace}
\usepackage{subfig}
\usepackage{xcolor}
\usepackage{hhline}
\usepackage{multirow}
\usepackage{soul}

\usepackage{array}
\usepackage{comment}
\usepackage{listings}
\usepackage{dsfont}
\usepackage{algorithmic}
\usepackage{xurl}
\usepackage{setspace}
\usepackage{graphicx}
\usepackage{caption}
\usepackage{ltablex}

\usepackage{ifluatex}

\usepackage{outlines}

\begin{document}

\newcommand*{\affmark}[1][*]{\textsuperscript{#1}}
\newcommand*{\affaddr}[1]{#1}

% \title{\LARGE \name: Fast Context Loading for Language Model Applications}
\title{\name: KV Cache Native Storage Hierarchy for Low-Delay and High-Quality Language Model Serving 
}
\author{
  \normalsize
  Shaoting Feng$^{*,1}$ \quad
  Hanchen Li$^{*,1}$ \quad
  Kuntai Du$^1$ \quad
  Zhuohan Gu$^1$ \quad
  Yuhan Liu$^1$ \quad
  Jiayi Yao$^1$ \quad
  Siddhant Ray$^1$ \quad
  Samuel Shen$^1$ \quad
  Yihua Cheng$^1$ \quad
  Ganesh Ananthanarayanan$^2$ \quad
  Junchen Jiang$^1$ \\
  \normalsize
  $^1$University of Chicago \hspace{.4in} $^2$Microsoft
}

\input{macros}

\input{sec-abstract}

\maketitle
\renewcommand{\thefootnote}{\fnsymbol{footnote}}
\footnotetext[1]{Equal contribution}
\renewcommand{\thefootnote}{\arabic{footnote}}

\pagestyle{plain}

\input{sec-intro}

\input{sec_policy_v1}
% \input{sec_design_v1}

\input{sec-eval-v2}

\bibliographystyle{plain}
\bibliography{citations} 

%%%%%%%%%%%%%%%%%%%%%%%%%%%%%%%%%%%%%%%%%%%%%%%%%%%%%%%%%%%%%%%%%%%%%%%%%%%%%%%%
\end{document}

%% file: macros.tex
%!TEX root = main.tex
%!TEX spellcheck = en_US
\newcommand{\yhedit}[1]{{\color{red} #1}}

\newcommand{\edit}[1]{{\color{black} #1}}
\newcommand{\shaoting}[1]{{\footnotesize\color{brown}(Shaoting: #1)}}
\newcommand{\jiayi}[1]{{\color{red}(Jiayi: #1)}}
\newcommand{\shanedit}[1]{{\color{red} #1}} 
\newcommand{\zgu}[1]{{\color{blue}(Joshua: #1)}}
\newcommand{\mm}[1]{{\color{violet}(Michael:#1)}}
\newcommand{\jc}[1]{{\footnotesize\color{blue}{(JC: #1)}}}
\newcommand{\qz}[1]{{\footnotesize\color{cyan!70!blue}{(Qizheng: #1)}}}
\newcommand{\kt}[1]{{\footnotesize\color{cyan!70!blue}{(Kuntai: #1)}}}
\newcommand{\jcedit}[1]{{\color{orange} #1}} 
\newcommand{\yh}[1]{{\footnotesize\color{deepgreen}{(Yuhan: #1)}}}
\newcommand{\todo}[1]{{\color{red}{(TODO: #1)}}}
\definecolor{darkkhaki}{rgb}{0.74, 0.72, 0.42}
\newcommand{\hc}[1]{{\color{darkkhaki}{(LHC: #1)}}}
\newcommand{\ga}[1]{{\color{cyan!70!blue}{(Ganesh: #1)}}}
\newcommand{\sid}[1]{{\footnotesize \color{brown!60!blue}{(Sid: #1)}}}
\definecolor{yccolor}{RGB}{240, 24, 240}
\newcommand{\yc}[1]{{\color{yccolor}{\footnotesize{[Yihua: #1]}}\xspace}}
% Changes here

% \newcommand{\yhedit}[1]{}

% \newcommand{\edit}[1]{}
% \newcommand{\shan}[1]{}
% \newcommand{\shanedit}[1]{} 
% \newcommand{\hank}[1]{}
% \newcommand{\mm}[1]{}
% % \newcommand{\cc}[1]{{\color{cyan!70!blue}{(Wan: #1)}}}
% \newcommand{\jc}[1]{}
% \newcommand{\jcedit}[1]{{\color{orange} #1}} 
% \newcommand{\yh}[1]{}
% % \newcommand{\kt}[1]{{\color{brown}{(Kuntai: #1)}}}
% \newcommand{\todo}[1]{}
% \definecolor{darkkhaki}{rgb}{0.74, 0.72, 0.42}
% \newcommand{\hc}[1]{}
% \newcommand{\ga}[1]{}
% \newcommand{\sid}[1]{}
% \definecolor{yccolor}{RGB}{240, 24, 240}
% \newcommand{\yc}[1]{}

% \newcommand{\shan}[1]{}
% \newcommand{\hank}[1]{ }
% \newcommand{\mm}[1]{ }
% \newcommand{\cc}[1]{ }
% \newcommand{\jc}[1]{ }
% \newcommand{\yh}[1]{ }
% \newcommand{\kt}[1]{ }

% \newcommand{\todo}[1]{{\color{red}{#1}}}

% \newcommand{\shan}[1]{}
% \newcommand{\hh}[1]{}
% \newcommand{\hank}[1]{}
% \newcommand{\mm}[1]{} 
% \newcommand{\cc}[1]{}
%  \newcommand{\jc}[1]{}
% \newcommand{\yh}[1]{}

\newcommand{\NumTokens}{\ensuremath{{N}}\xspace}
\newcommand{\NumChannels}{\ensuremath{{c}}\xspace}
\newcommand{\NumLayers}{\ensuremath{{l}}\xspace}
\newcommand{\Level}{\ensuremath{{k}}\xspace}

\newcommand{\Decision}{Decision\xspace}
\newcommand{\x}{\ensuremath{\mathbf{x}}\xspace}
\newcommand{\y}{\ensuremath{\mathbf{y}}\xspace}
\newcommand{\z}{\ensuremath{z}\xspace}
\newcommand{\X}{\ensuremath{\mathbf{X}}\xspace}
\newcommand{\TestNum}{\ensuremath{T}\xspace}
\newcommand{\APIcomp}{App \circ Filter\xspace}
\newcommand{\Filter}{F\xspace}
\newcommand{\DNN}{DNN\xspace}
\newcommand{\API}{API\xspace}
\newcommand{\Appdecision}{App\xspace}
\newcommand{\total}{\ensuremath{t}\xspace}
\newcommand{\NTarget}{\ensuremath{N}\xspace}
\newcommand{\other}{\ensuremath{O}\xspace}

\newcommand{\Rate}{IDR\xspace}

\newcommand{\T}{\ensuremath{T}\xspace}
\newcommand{\Match}{\ensuremath{M}\xspace}
\newcommand{\BeforeBreak}{\ensuremath{B}\xspace}

\newcommand{\App}{$A$\xspace}
\newcommand{\M}{\ensuremath{M}\xspace}
\newcommand{\F}{\ensuremath{\alpha}\xspace}
\newcommand{\G}{\ensuremath{\beta}\xspace}
\newcommand{\Q}{\ensuremath{\gamma}\xspace}
\newcommand{\ClassId}{\ensuremath{c}\xspace}
\newcommand{\LabelId}{\ensuremath{l}\xspace}
\newcommand{\InputId}{\ensuremath{i}\xspace}
\newcommand{\TraverseId}{\ensuremath{j}\xspace}
\newcommand{\score}{effective score\xspace}
\newcommand{\scores}{effective scores\xspace}
\newcommand{\Tg}{\ensuremath{TC}\xspace}
\newcommand{\TargetSet}{\ensuremath{G}\xspace}
\newcommand{\MaxScore}{\ensuremath{f(\y)}\xspace}

\newcommand{\summary}{decision-process summary\xspace}
\newcommand{\Summary}{Decision-process summary\xspace}
\newcommand{\summaries}{decision-process summaries\xspace}
\newcommand{\target}{target class\xspace}
\newcommand{\targets}{target classes\xspace}
\newcommand{\Target}{Target class\xspace}
\newcommand{\process}{software decision process\xspace}
\newcommand{\Process}{Software decision process\xspace}
\newcommand{\bestapi}{Best-of-all API*\xspace}
\newcommand{\qa}{interactive Q \& A\xspace}
\newcommand{\rag}{retrieval-augmented generation\xspace}
\newcommand{\kv}{key-value cache\xspace}
\newcommand{\scis}{Scissorhands\xspace}
\newcommand{\recomp}{Loading text context\xspace}

\newcommand{\name}{AdaptCache\xspace}
\newcommand{\modeld}{$\textrm{ChameleonAPI}_{basic}$\xspace}
\newcommand{\tool}{ChameleonAPI\xspace}
\newcommand{\toolbasic}{\ensuremath{\textrm{ChameleonAPI}_{basic}}\xspace}
\newcommand{\code}[1]{{\texttt{#1}}}
\newcommand{\term}[1]{\textsf{#1}}
\newcommand{\wl}{\emph{whitelist}\xspace}
\newcommand{\wls}{\emph{whitelists}\xspace}
\newcommand{\ift}{\emph{if/then branch}\xspace}
\newcommand{\ifts}{\emph{if/then branches}\xspace}

\newcommand{\fillme}{{\bf XXX}\xspace}

\newcommand*\circled[1]{\tikz[baseline=(char.base)]{
            \node[shape=circle,fill,inner sep=2pt] (char) {\textcolor{white}{\footnotesize{#1}}};}}

\newcounter{packednmbr}
\newenvironment{packedenumerate}{\begin{list}{\thepackednmbr.}{\usecounter{packednmbr}\setlength{\itemsep}{0.5pt}\addtolength{\labelwidth}{-4pt}\setlength{\leftmargin}{2ex}\setlength{\listparindent}{\parindent}\setlength{\parsep}{1pt}\setlength{\topsep}{0pt}}}{\end{list}}
\newenvironment{packeditemize}{\begin{list}{$\bullet$}{\setlength{\itemsep}{0.5pt}\addtolength{\labelwidth}{-4pt}\setlength{\leftmargin}{2ex}\setlength{\listparindent}{\parindent}\setlength{\parsep}{1pt}\setlength{\topsep}{2pt}}}{\end{list}}
\newenvironment{packedpackeditemize}{\begin{list}{$\bullet$}{\setlength{\itemsep}{0.5pt}\addtolength{\labelwidth}{-4pt}\setlength{\leftmargin}{\labelwidth}\setlength{\listparindent}{\parindent}\setlength{\parsep}{1pt}\setlength{\topsep}{0pt}}}{\end{list}}
\newenvironment{packedtrivlist}{\begin{list}{\setlength{\itemsep}{0.2pt}\addtolength{\labelwidth}{-4pt}\setlength{\leftmargin}{\labelwidth}\setlength{\listparindent}{\parindent}\setlength{\parsep}{1pt}\setlength{\topsep}{0pt}}}{\end{list}}
\let\enumerate\packedenumerate
\let\endenumerate\endpackedenumerate
\let\itemize\packeditemize
\let\enditemize\endpackeditemize

\newcommand{\tightcaption}[1]{\vspace{-0.15cm}\caption{{\normalfont{\textit{{#1}}}}}\vspace{-0.3cm}}
\newcommand{\tightsection}[1]{\vspace{-0.1cm}\section{#1}\vspace{-0.00cm}}
\newcommand{\tightsectionstar}[1]{\vspace{-0.17cm}\section*{#1}\vspace{-0.08cm}}
\newcommand{\tightsubsection}[1]{\vspace{-0.1cm}\subsection{#1}\vspace{-0.02cm}}
\newcommand{\extightsubsection}[1]{\vspace{-0.3cm}\subsection{#1}\vspace{-0.02cm}}
\newcommand{\tightsubsubsection}[1]{\vspace{-0.01in}\subsubsection{#1}\vspace{-0.01cm}}

\newcommand{\eg}{{\it e.g.,}\xspace}
\newcommand{\ie}{{\it i.e.,}\xspace}
\newcommand{\etal}{{\it et.~al}\xspace}
\newcommand{\bigO}{\mathrm{O}}
\newcommand{\twlog}{w.l.o.g.\xspac}

\newcommand{\myparashort}[1]{\vspace{0.05cm}\noindent{\bf {#1}}~}
\newcommand{\mypara}[1]{\vspace{0.05cm}\noindent{\bf {#1}:}~}
\newcommand{\myparatight}[1]{\vspace{0.02cm}\noindent{\bf {#1}:}~}
\newcommand{\myparaq}[1]{\vspace{0.05cm}\noindent{\bf {#1}?}~}
\newcommand{\myparaittight}[1]{\smallskip\noindent{\emph {#1}:}~}
\newcommand{\question}[1]{\smallskip\noindent{\emph{Q:~#1}}\smallskip}
\newcommand{\myparaqtight}[1]{\smallskip\noindent{\bf {#1}}~}

\newcommand{\cmark}{\ding{51}}%
\newcommand{\xmark}{\ding{55}}%

% \definecolor{codegreen}{rgb}{0,0.6,0}
% \definecolor{codegray}{rgb}{0.5,0.5,0.5}
% \definecolor{codepurple}{rgb}{0.58,0,0.82}
% \definecolor{backcolour}{rgb}{0.95,0.95,0.92}

% \lstdefinestyle{overleaf_style}{
%     backgroundcolor=\color{backcolour},   
%     commentstyle=\color{codegreen},
%     keywordstyle=\color{magenta},
%     numberstyle=\tiny\color{codegray},
%     stringstyle=\color{codepurple},
%     basicstyle=\ttfamily\footnotesize,
%     breakatwhitespace=false,         
%     breaklines=true,                 
%     captionpos=b,                    
%     keepspaces=true,                 
%     numbers=left,                    
%     numbersep=5pt,                  
%     showspaces=false,                
%     showstringspaces=false,
%     showtabs=false,                  
%     tabsize=2
% }

\definecolor{backcolour}{rgb}{0.96,0.96,0.96}
\definecolor{codegray}{rgb}{0.5,0.5,0.5}
\definecolor{deepblue}{rgb}{0,0,0.6}
\definecolor{deepred}{rgb}{0.6,0,0}
\definecolor{deepgreen}{rgb}{0,0.5,0}
\lstdefinestyle{mystyle}{
    backgroundcolor=\color{backcolour},   
    commentstyle=\color{codegreen},
    morekeywords={self, True},
    keywordstyle=\color{deepblue},
    numberstyle=\tiny\color{codegray},
    emph={MyClass,__init__,EncodingType,Image},
    emphstyle=\color{deepred},
    stringstyle=\color{deepgreen},
    basicstyle=\ttfamily\footnotesize,
    breakatwhitespace=false,         
    breaklines=true,                 
    captionpos=b,                    
    keepspaces=true,                 
    numbers=left,                    
    numbersep=5pt,                  
    showspaces=false,                
    showstringspaces=false,
    showtabs=false,                  
    tabsize=1
}

\newcommand{\mycaption}[3]{
\begin{spacing}{0.95}
\caption{
\label{#1}
{\bf #2. }
{\it \small #3}
}
\end{spacing}
}

\newcommand{\vminfive}{\vspace{-5pt}}
\newcommand{\vminten}{\vspace{-10pt}}
\newcommand{\vminfifteen}{\vspace{-15pt}}
\newcommand{\vmintwenty}{\vspace{-20pt}}
\newcommand{\vmintwentyfive}{\vspace{-25pt}}

\def \hmina {\hspace{-0.1in}}
\def \hminb {\hspace{-0.2in}}

%% file: sec-abstract.tex
%!TEX root = main.tex
%!TEX spellcheck = en_US

\begin{abstract}

Large language model (LLM) applications often reuse previously processed context, such as chat history and documents, which introduces significant redundant computation.
Existing LLM serving systems address such redundant computation by storing the KV caches of processed context and loading the corresponding KV cache when a new request reuses the context.
Further, as these LLM applications scale, the total size of KV caches becomes excessively large and requires both DRAM and SSD for full storage.

However, prior work that stores KV caches in DRAM and SSD suffers from high loading delays, as most KV cache hits come from SSD, which is slow to load.
To increase the KV cache hit rate on DRAM, we identify lossy KV cache compression as a promising approach. We design a lossy compression system that decides the compression algorithm, compression rate and device placement for each KV cache entry to maximise DRAM hits and minimise loading delay without significantly degrading generation quality.
Compared to various static compression baselines across three tasks, our system \name achieves 1.43--2.4 $\times$ delay savings at the same quality and 6--55\% quality improvements at the same delay.

\end{abstract}

%% file: sec-intro.tex
%!TEX root = main.tex
%!TEX spellcheck = en_US

\tightsection{Introduction}

With the accelerating adoption of Large Language Models (LLMs), inference systems must provide low-delay responses to meet growing user demands. To achieve this, many systems store and reuse the KV cache of previously processed contexts (\eg chat history) to reduce redundant computation and speed up inference. 
However, due to the rapid growth of LLM applications, the size of KV caches also quickly increases and exceeds the storage capacity of both GPU and CPU memory. 
% \kt{We may need some numbers here. Example: for models like DeepSeek-R1, one H100 node with 80GB GPU memory and 150GB CPU memory can only save the KV caches of 100K tokens, which roughly translates to the chatting history of 50 users.}\shaoting{50 long sessions}
For example, even with two H100 nodes (each with 80GB GPU memory) and a total of 150GB CPU memory, models like \textit{Llama-3.1-70B-Instruct} can only store the KV caches of 500K tokens, roughly corresponding to the chat histories of 30 users.
This issue becomes more severe in agentic applications, where even a single run involves multiple agents that continuously exchange the text with each other, resulting in a context of millions of tokens and the corresponding KV cache size of 3TB.
% they need to deal with more and more users (and their corresponding contexts), 
% the amount of KV cache that needs to be stored (\eg the chatting history of each user) also grow correspondingly significantly grows these days, due to the rapid growth of chatting applications as popular inference applications scale up (e.g., ChatGPT processes over 2.5 billion prompts per day \cite{roth2025chatgpt}),
This underscores the need to fully leverage hierarchical storage within the datacenter to ensure high cache hit rates, reduce delay, and increase throughput.

There is a line of work~\cite{chen2025impress,gao2024cost} that explores storing KV caches in hierarchical storage (in this paper we focus on DRAM and SSD).
However, we observe that these approaches still have high delay, because high-speed devices have limited storage space, which forces most of the KV caches to be stored to SSD, leading to slow KV cache loading time overall.
% This significantly increases the loading time of these KV cache entries.
To allow more cache hits on DRAM, we identify lossy KV cache compression \cite{liu2024kivi,xiao2023efficient} as a promising approach: by lossily compressing the KV caches, we can store much more KV cache entries in CPU memory without significant degredation of generation quality\footnote{Generation quality is defined as the similarity between the generated answer after compression and the original prefill answer, measured by task-specific metrics, \ie F1, ROUGE-L \cite{lin-2004-rouge} or CodeBLEU \cite{ren2020codebleumethodautomaticevaluation}.}.
% which means that we can store much more KV caches in CPu
% For example, \kt{An example. Maybe a method that can quantize KV caches to 2 bits and still not change the summarization results, something like that}.
% the we can quantize KV Cache from 16 bits to 2 bits or drop tokens with relatively low attention scores. 
% With the size compressed to one-eighth of its original value, the high-speed device can store significantly more KV cache, increasing the cache hit rate and reducing delay.

As a result, our goal is to design a lossy KV cache storage system on a hierarchical storage (DRAM and SSD) that ensures low inference latency with minimum quality degredation.

\begin{figure}[t]
\centering
    \includegraphics[width=0.99\columnwidth]{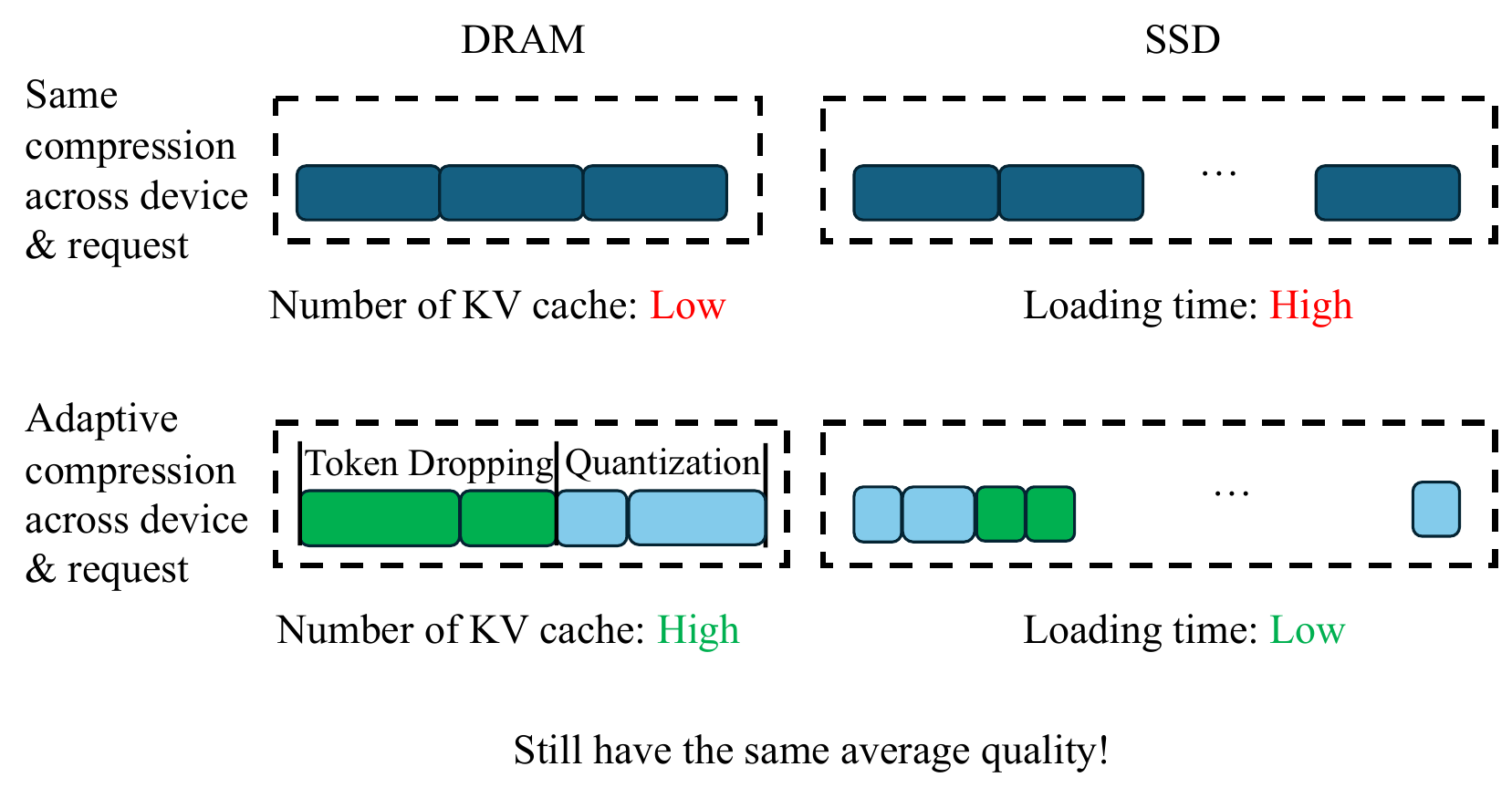}
    \tightcaption{\name increases cache hit in high-speed device and reduces loading time in low-speed device while maintaining high generation quality}
    \label{fig:e2e_fig_only_hit2}
\end{figure}

This paper argues that, instead of applying the same compression (\ie using the same compression algorithm and compression rate across devices) to all KV caches, we need to adaptively adjust the compression algorithm, compression rate and device placement to achieve optimal KV cache compression.
Concretely:
\begin{packeditemize}
    \item Compression algorithm needs to be aware of the context - Some texts are only important at the beginning and the end (which is suitable for token dropping), but some texts aim to provide new information for the LLM (where quantization might be better).
    \item Compression ratio and device placement need to be aware of the content and the reuse frequency. Some KV caches are frequently reused and easy-to-be-compressed and they need to be stored in CPU memory with a high compression ratio, but some KV caches are not reused frequently and difficult to compress, so it may be suitable to place them on disk with a low compression ratio.
\end{packeditemize}
Figure~\ref{fig:e2e_fig_only_hit2} shows that, if we optimally adapt the compression algorithm, rate and device, the potential of improvement is huge: we can achieve much higher KV cache hit rate on CPU and much lower average loading time from disk, while still having comparable accuracy as applying same compression across devices.

% for each request can increase the cache hit rate and reduce delay while maintaining high quality.

%The ideal compression choice to achieve a favourable quality-delay trade-off should adapt to storage device heterogeneity and various contents.

% Further, lossy compression also makes it more challenging to perform KV cache eviction when the storage is full
% for the same context, no KV cache copy is obviously worse than the others: one may have higher loading delay but offer better accuracy, while others load faster but suffer from lower accuracy.

\begin{figure*}[t]
\centering
    \includegraphics[width=1.8\columnwidth]{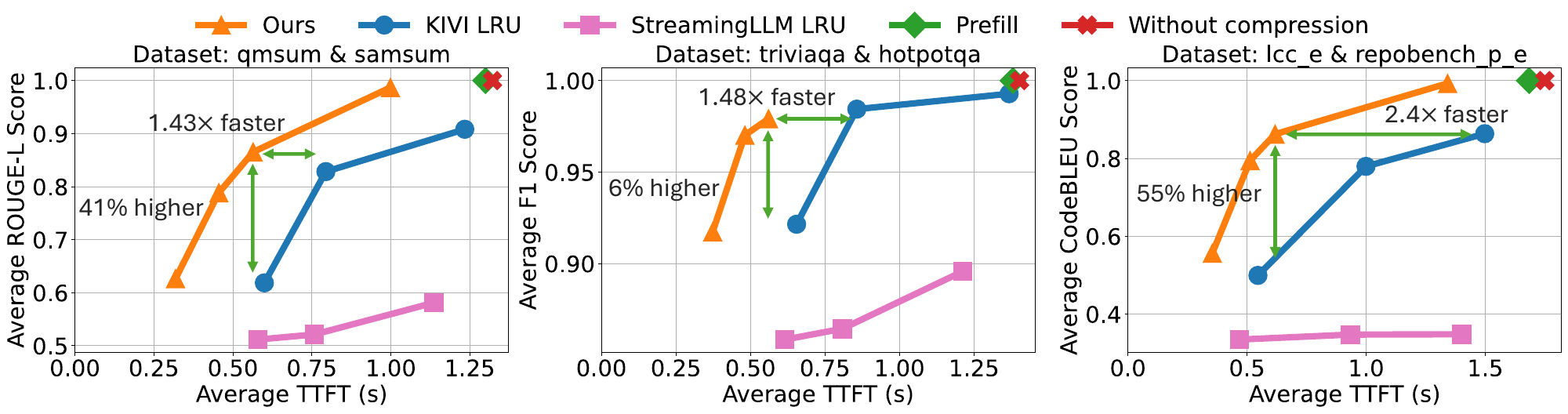}
    \tightcaption{\name achieves 1.43--2.4~$\times$ lower TTFT and 6--55\% higher quality compared to fixed compression method and rate}
    \label{fig:e2e_fig_only_hit}
\end{figure*}

To address these challenges, we propose \name, the first hierarchical KV cache storage system leveraging lossy KV cache compression.
The core of \name is a utility metric that allows us to quantify the effect of storing KV caches in terms of both quality and loading delay.
Then, we use \emph{marginal utility gain} to evaluate the effect of each design decisions -- compression algorithm, compression rate, cache placement and eviction -- and pick the design decision with the highest marginal utility gain.
Acknowledging that this design is greedy and not necessarily optimal, we argue that it suffices to deliver significant delay reduction and accuracy improvement in practice, and optimal solution is not trackable since it is NP-hard.

To evaluate \name, we use the popular \textit{Llama-3.1-8B-Instruct} model and three LongBench datasets.
Compared to the strongest fixed compression baseline, \name is 1.43--2.4$\times$ faster at the same quality and improves quality by 6--55\% under the same delay.

%% file: sec_policy_v1.tex
\tightsection{\name Design}

\name has three main components: an estimator, a policy optimizer, and an executor. The estimator does offline profiling to estimate the quality–delay curve for each entry, compression method and storage device. The policy optimizer makes compression decisions for incoming and existing KV cache entries. The executor implements the compression decisions.
% implements the selected compression algorithms and does offloading.

\mypara{Estimator} Our offline profiler uses dummy questions to measure the transfer delays of each storage device and decompression overhead. It also samples ten entries from each dataset, using questions generated by GPT 4o, to generate the quality–compression rate curve for each compression method. We estimate the future cache hit frequency of one KV cache entry using its historical hit frequency.

\noindent \textbf{Utility definition:} 
The objective of \name is to minimize average system delay while maintaining high quality. To capture this trade-off, we compute a quality-delay weighted sum for each KV cache entry. 
Since some entries are reused more frequently than others, storing these in limited storage can increase the overall hit rate thus reducing delay. Therefore, we define each entry's utility as the product of its estimated future occurrence frequency and its quality-delay weighted sum:

\vspace{-0.2cm}
{
\footnotesize
\[
\mathit{Utility}(i)=\mathit{Freq}(i) \cdot ( 
        \alpha \times \mathit{Quality}(i, M_i, R_i)
        - \frac{\mathit{size}(i, M_i, R_i)}{\mathit{Bandwidth}}
\]
}
\noindent where $i$ represents the i-th KV cache entry, $M_i$ and $R_i$ are the compression method and rate used.

The objective of \name can be expressed as to maximise the total utility across all KV cache entries.
% \[
% \begin{aligned}
%     \text{Maximize} \quad & \sum_{d} \sum_{i\in d} \mathit{Utility} \\
%     \text{with} \quad
%         & \sum \mathit{size}(i, M_i, R_i) \leq \mathit{StorageSize},
% \end{aligned}
% \]
% where $i$ represents the i-th KV cache entry, $M_i$ and $R_i$ are the compression method and rate used.
% By measuring utility drop per unit of space, we effectively balance the relative importance of each entry against the limited storage available.
Mathematically, this is an NP-hard Multi-Choice Knapsack Problem (MCKP). We adopt a greedy approach based on the textbook solution~\cite{kellerer2004multidimensional} that achieves Linear Programming Optimality. When a new KV cache entry is generated, 
% we logically add it to storage. Then for each KV cache entry, we compute:
we compute the following metric for both the new entry and all entries in storage. We define

\vspace{-0.2cm}
{
\footnotesize{
\[
 \frac{\mathit{Utility}(i,m) - \mathit{Utility}(i,n)}{\mathit{size}(i) \times (m - n)}
\]
}
}
\noindent
as the \textbf{marginal utility drop}, \ie the utility drop per unit of space saved if we compress furthermore or evict the entry. $m$ and $n$ are different compression rates.

For each storage tier, we will greedily decide the compression choice (to evict, store in full quality or compress each entry) that results in the minimal marginal utility drop.

% The key to our policy is the metric for our greedy decision: the utility drop per unit of space. This metric reflects both the storage device heterogeneity and the KV cache heterogeneity. The delay estimation take into account the storage device speed and KV cache length. The quality estimation take into account the KV cache content's sensitivity to different compression method and robustness to different compression rates. THe per unit of space take into account the KV cache length.

% % \mypara{Executor} We introduce a temporary main memory buffer to store new KV cache entries and those requiring further compression. We use customised CUDA kernels to offload KV cache to main memory and perform non-blocking offloading to SSD using LMCache \cite{lmcache}.

%% file: sec-eval-v2.tex
\tightsection{Preliminary Results}

\mypara{Evaluation setup} 
%We implement \name on top of vLLM \cite{kwon2023efficient} and LMCache \cite{lmcache} using about 1,000 lines of Python code. 
We use 1,100 contexts from six LongBench datasets \cite{bai2023longbench,gliwa2019samsum,zhong2021qmsum,joshi2017triviaqa,yang2018hotpotqa,liu2023repobench,guo2023longcoder} to evaluate \name, covering three task types, summarization, question answering (QA) and coding.
% , with datasets combined to simulate heterogeneous workloads.
These datasets do not contain the timestamps for request arrival, thus we follow previous works to generate request arritval timestamps using Poisson distribution with different request rates~\cite{kwon2023efficient,xue2020mt5,gao2024cost}. 
Experiments are run on one NVIDIA A100 GPU (100 GB DRAM, 400 GB SSD) using \textit{Llama-3.1-8B-Instruct} model.
%, ensuring no queued requests due to sufficient prefilling and decoding throughput\yh{why do we need to ensure no queued requests? }. 
The disk reading throughput is 1 GB/s. 
The baselines we compare are: 
\begin{packeditemize}
    \item \emph{Without Compression} denotes offloading KV cache to DRAM and SSD without compression.
    \item \emph{KIVI LRU} / \emph{StreamingLLM LRU} use KIVI~\cite{liu2024kivi} and StreamingLLM~\cite{xiao2023efficient} with fixed compression rates, and offload KV cache to DRAM and SSD using a Least Recently Used (LRU) policy.
    % \item \emph{StreamingLLM LRU} uses StreamingLLM \cite{?} with fixed compression rates then offloads KV cache to main memory and SSD.
    \item \emph{Prefill} denotes prefilling (recomputation).
\end{packeditemize}
 
% \mypara{TTFT reduction} As shown in Figure~\ref{fig:e2e_fig_only_hit}, \name significantly reduces TTFT across three tasks. With an SSD bandwidth of 1 GB/s, \name achieves a 50\% reduction in TTFT compared to LRU static compression using KIVI. Against pure offloading and prefilling, and with a quality drop SLO of 5\%, \name reduces TTFT by 20\% for summarisation, 69.6\% for QA, and 35.7\% for coding.

% \mypara{Quality gains} The same figure also illustrates \name's strong ability to maintain output quality across all three tasks. When the delay SLO is set to half that of prefilling or pure offloading, \name limits the quality loss to 11\%, 1\%, and 10\% for summarisation, QA, and coding, respectively, while static compression results in higher losses of 35\%, 7\%, and 35\%.
\mypara{Overall results} As shown in Figure~\ref{fig:e2e_fig_only_hit}, across three tasks, \name consistently reduces time to first token (TTFT) compared to all baselines. Compared to naive prefill and offloading, \name reduces TTFT by 56\%, achieving within 15\% quality drop. 
Compared to KIVI, \name reduces TTFT by 69\% at the same quality. Finally, compared to StreamingLLM, \name greatly improves quality by 15--89\% at the same TTFT. 

\mypara{Understanding \name's improvements}
\name outperforms approaches that use no compression or fixed compression rates by detecting speed differences between devices and aggressively compressing the KV cache. This strategy significantly increases the hit rate on high-speed devices, leading to much lower loading delay and reduced TTFT. For example, in the coding task, KIVI LRU achieves a DRAM cache hit rate of 38\% when quantized to 2 bits. In contrast, our method achieves hit rates of 81\%, 56\%, 44\%, and 11\%, for different weights between delay and quality.
% On slower devices, compression further helps by reducing the amount of data that needs to be transferred. 
% \name also prioritizes compressing and storing large KV caches in DRAM, while for short requests, it opts to recompute, ensuring good quality with minimal delay,
%or offload without compression to minimize overall compression overhead.
% \name priotitzes compressing KV cache entries with low sensitivity to compression (low quality drop under same compression rate). \name chooses appropriate compression method that has least quality drop.
\name also prioritises compressing and storing KV cache from longer contexts or those with high information redundancy, and selects optimal compression algorithms for each entry, thereby ensuring overall high response quality.